\documentclass[12pt,a4paper,dvips,epsfig]{article}
\usepackage{a4p}
\usepackage{cite,mcite}
\usepackage{epsfig}   
\usepackage{physics}
\usepackage{l3_title,ifthen,Lep}
\journalname{Phys. Lett. B}
\date{July 17, 2001}
\preprint{2001-054}
\Lep{1}
\newlength{\capindent}
\newlength{\capwidth}
\newlength{\figwidth}
\newcommand{\icaption}[2][!*!,!]{\hspace*{\capindent}%
  \begin{minipage}{\capwidth}
    \ifthenelse{\equal{#1}{!*!,!}}%
      {\caption{#2}}%
      {\caption[#1]{#2}}
  \end{minipage}}
\newcommand{\pho}{\phantom{0}}

\newcommand{\eee}{\mbox{${\rm e}^{+}{\rm e}^{-} \rightarrow {\rm e}^{+} {\rm e}^{-}(\gamma)$}}
\newcommand{\emm}{\mbox{${\rm e}^{+}{\rm e}^{-} \rightarrow {\mu  }^{+} {\mu  }^{-}(\gamma)$}}

\newcommand{\taupg}{\mbox{${\rm e}^{+} {\rm e}^{-}\rightarrow\tau^{+}\tau^{-}(\gamma)$}}
\newcommand{\boldtaupg}{\mbox{\boldmath${\mathrm e}^{+} {\mathrm e}^{-}\rightarrow\tau^{+}\tau^{-}(\gamma)$}}

\newcommand{\taone}{\mbox{$\tau^-\rightarrow {\mathrm{a_1^{\kern -0.1em -}}}\nu_{\tau}$}}

\def\br {\ensuremath{{\cal{B}}}}
\newcommand{\brt}[1]{\br(\tau\rightarrow (#1-prong))}
\newcommand{\brts}[1]{\br(#1-prong)}
\begin{document}

\begin{titlepage}

\mathversion{bold}
\title{Measurement of the  Topological Branching Fractions of the $\tau$
lepton at LEP}

\author{The L3 Collaboration}

\mathversion{normal}

%
% The abstract
%
\begin{abstract}

Using  data collected with the L3 detector at LEP from 1992 to 1995
on the Z peak, we determine the branching fractions of the $\tau$
lepton
into one, three and five charged particles to be: 
\begin{eqnarray*}
 & \brt{1} &  
                      = 85.274 \pm 0.105 \pm 0.073 \% , \\
 & \brt{3} &
                      =  14.556 \pm 0.105 \pm 0.076 \%, \\  
 & \brt{5} &
                      = \pho  0.170 \pm 0.022 \pm 0.026 \%.  
\end{eqnarray*}
\noindent
The first uncertainties are statistical and the second systematic.
The accuracy of these measurements alone
is similar to that of the current world average.

\end{abstract}

%
% Adds "To be submitted to ..." or "Submitted to ...", if relevant
%
\submitted
\end{titlepage}

\noindent
%%%%%%%%%%%%%%%%%%%%%%%%%%%%%%%%%%%%%%%%%%%%%%%%%%%%%%%%%%%%%%%%%%%%%%%%%%%%%%%
% Introduction
%%%%%%%%%%%%%%%%%%%%%%%%%%%%%%%%%%%%%%%%%%%%%%%%%%%%%%%%%%%%%%%%%%%%%%%%%%%%%%%
\section*{Introduction}
Measurements of the topological branching fractions of the $\tau$ lepton 
and the sum of measurements of the exclusive branching fractions 
were previously inconsistent. 
Solving this ``one-prong puzzle'' motivated many precise 
determinations of the exclusive $\tau$ branching fractions at the permille 
level~\cite{PDG2000} but only a few less precise determinations of the topological branching
fractions~\cite{argus_92D} have been performed.

In this letter, we present a new 
measurement of
the topological branching
fractions using data  
collected by the L3 detector at LEP on the Z resonance.
These results supersede those of our previous publication~\cite{l3_91F}.
Here we follow the convention that tracks stemming from neutral kaon decays
are not accounted for in the topology.  
Another  measurement of the topological $\tau$ branching fractions
%from the complete data set on the Z resonance 
was recently reported in Reference~\citen{delphip}.

The measurement entails a selection of \taupg~ events followed by 
an event topology reconstruction, which must be precisely understood.
The reconstructed topology is influenced by photon 
conversions, subdetector inefficiencies and resolution limitations. 
Detailed studies of these effects are performed in order to control the systematic
uncertainties to the level of the statistical uncertainties.

%%%%%%%%%%%%%%%%%%%%%%%%%%%%%%%%%%%%%%%%%%%%%%%%%%%%%%%%%%%%%%%%%%%%%%%%%%%%%%%
% Data and Monte Carlo samples
%%%%%%%%%%%%%%%%%%%%%%%%%%%%%%%%%%%%%%%%%%%%%%%%%%%%%%%%%%%%%%%%%%%%%%%%%%%%%%%
\section*{Data and Monte Carlo samples}

The data used  were collected with the L3 detector~\cite{l3_det}
at LEP from 1992 to 1995
on the Z peak, corresponding to an integrated luminosity of 
92.6 pb$^{-1}$.
The most crucial subdetectors for this analysis are: the central tracking system 
consisting of a silicon microvertex detector (SMD), 
a time expansion chamber (TEC) and proportional chambers measuring the Z coordinate,
the electromagnetic calorimeter composed of 
Bismuth Germanium Oxide (BGO) crystals, the hadron calorimeter
(HCAL) and the muon spectrometer. Detailed studies of the efficiencies of these
subdetectors using control samples are performed, yielding precise determinations
of the efficiencies.

For efficiency studies, 
\taupg~events are 
generated with the  KORALZ Monte Carlo generator~\cite{koralz1}.
Background estimations
are performed using the following 
 Monte Carlo generators:
%\noindent
KORALZ for \emm; BHAGENE~\cite{bhagene1} for 
\eee; JETSET~\cite{my_jetset}
 for ${\rm e^+e^-} \rightarrow {\rm q\bar{q}} (\gamma)$;
DIAG36~\cite{diag36}
 for ${\rm e^+e^-} \rightarrow {\rm e^+e^-}\ell^+\ell^-$, 
where $\ell={\rm e}$, $\mu$, or 
$\tau $.
   The Monte Carlo events are simulated in the L3 detector 
using the GEANT program~\cite{my_geant},
 which takes into account the effects
of energy loss, multiple scattering and showering.
Furthermore, time dependent detector inefficiencies are
considered.
These events are reconstructed with the same program as the one used for the
data.

%%%%%%%%%%%%%%%%%%%%%%%%%%%%%%%%%%%%%%%%%%%%%%%%%%%%%%%%%%%%%%%%%%%%%%%%%%%%%%%
% Detector Calibrations
%%%%%%%%%%%%%%%%%%%%%%%%%%%%%%%%%%%%%%%%%%%%%%%%%%%%%%%%%%%%%%%%%%%%%%%%%%%%%%%
\section*{Subdetector  efficiencies and calibrations}

Efficiency studies
of the subdetectors are done separately 
for each year of data taking.
As the year-by-year efficiency variations are small, 
average values are given in the following.

The efficiency of the TEC
to measure a track 
is studied using 
data samples of Bhabha and dimuon events 
and muons originating from $\tau$ decays.
The Bhabha and dimuon samples are selected
by requiring two energy deposits in the BGO or
two tracks in the muon spectrometer of about
the beam energy and back-to-back topology.
The muons in $\tau$ decays are identified as tracks in the 
muon chambers
pointing to the interaction region. In addition, the energy
deposits in the BGO and HCAL must be consistent 
with the expectation for a minimum ionising
particle (MIP). 

\begin{figure}[p]
\begin{center}
\mbox {\epsfig{figure=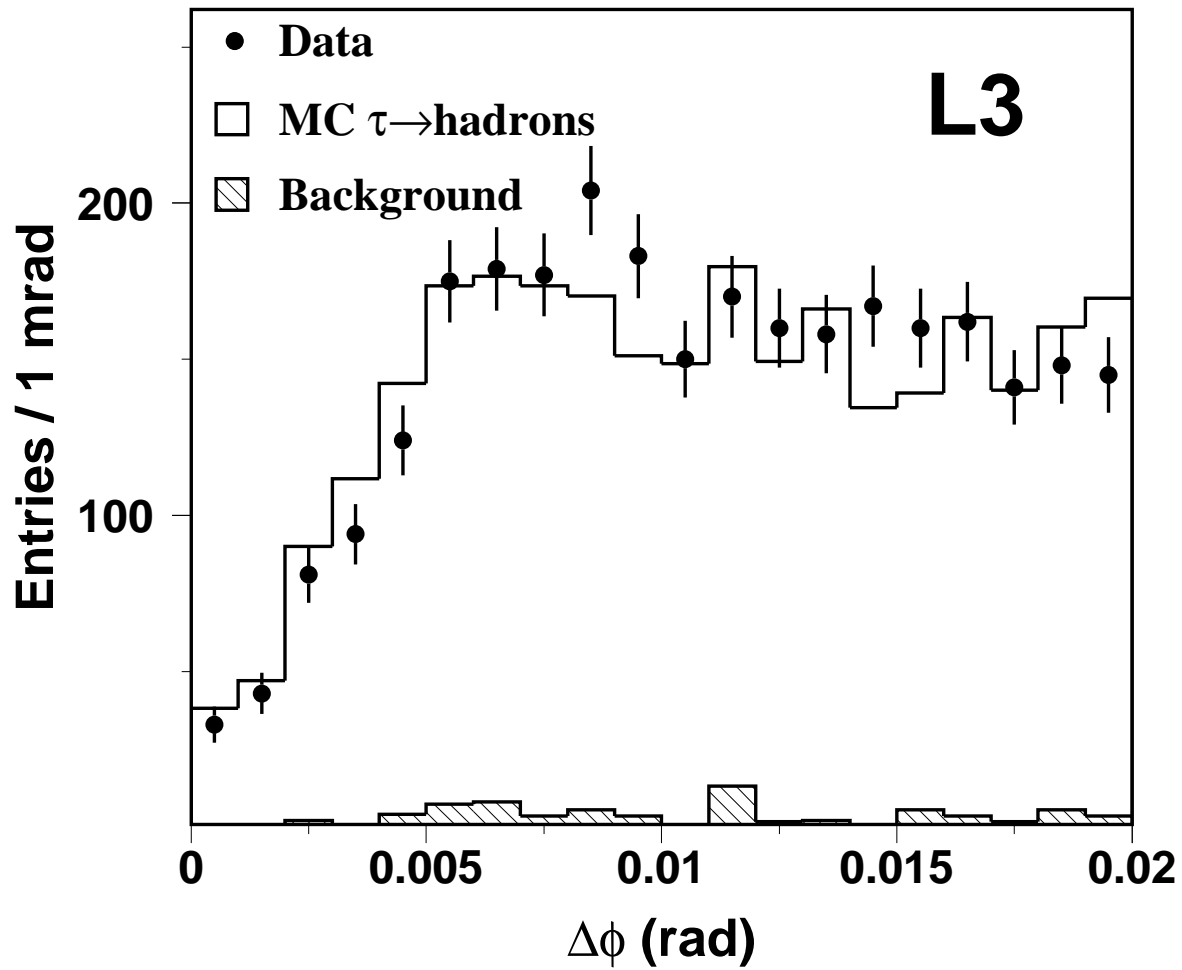,width=0.92\textwidth}}
\end{center}
\caption[]{\label{fig:dtr} Distribution of the 
azimuthal angle between two adjacent tracks $\Delta \phi$, where the
tracks must have a transverse momentum larger than $10 \GeV$.
}
\end{figure}

A track in the TEC must have at least 25 out of the 62 possible
hits, one or more hits in the innermost part of the chamber and to
span over more than 40 anode wires radially.
Its transverse momentum, $p_T$, must be larger than 2 \GeV. 
After rejecting tracks in the low resolution region adjacent to the
anode, the track finding
efficiency is found to be about 96\%,
almost independent of the track momentum.
The double track resolution of the TEC 
is determined from data
~\cite{L3TEC}
to be about 500$\mu$m and is modeled correspondingly
in the detector simulation.
As a cross check, the distributions of the azimuthal angle
between two adjacent tracks $\Delta \phi$, from data and 
Monte Carlo are compared in Figure~\ref{fig:dtr} for
small values of $ \Delta \phi$ and high momentum tracks
from 3-prong $\tau$ decays.
For $ \Delta \phi$ larger than
0.005\,rad excellent agreement is found.
The small discrepancy below 0.005\,rad
is taken into account as a systematic uncertainty.
 
The efficiency of the BGO to
detect an electromagnetically showering particle 
is determined using Bhabha
events to be about
99.5\%.
This efficiency is
found to be almost independent of the shower energy
from studies using ${\rm e^+e^-} \rightarrow {\rm e^+e^-e^+e^-}$
events.
In order
to estimate the efficiency of the BGO to detect a MIP,
$\tau$ decays into muons are used. A track in the 
muon spectrometer
is required, which points to the interaction region and
matches an energy deposit in the HCAL that
corresponds to a MIP. 
From these muons $ 97\%$  
induce a signal in the BGO. 
%The uncertainty is $0.3\%$ 

The same technique was used to estimate the HCAL and muon 
spectrometer efficiencies.
Using muons with a
track in the TEC, a MIP signal in the BGO and a matched 
muon spectrometer track, the efficiency of the HCAL
to detect such a particle is about 89\%.
% with an uncertainty of 0.3\%.
The muon spectrometer efficiency is found to be 74\% 
%with an uncertainty of 0.5\% 
using
$\tau$ decays with a track
in the TEC and a MIP signature in the BGO and  the HCAL. 

The subdetector efficiencies obtained 
from each year are used to correct the 
Monte Carlo simulation of the detector response
for the \taupg~and background processes.
The energy scales of the subdetectors are calibrated using
control data samples~\cite{l3pol}. The momentum scale of the central tracker
is verified to 0.5\% from 1 to $45\GeV$. The BGO and the muon spectrometer
scale uncertainties are 0.5\% at low energies and 0.05\% at high energy.
The scale uncertainty of the HCAL is 1\%.

%%%%%%%%%%%%%%%%%%%%%%%%%%%%%%%%%%%%%%%%%%%%%%%%%%%%%%%%%%%%%%%%%%%%%%%%%%%%%%%
% Backlash and $\gamma$ conversion track rejection
%%%%%%%%%%%%%%%%%%%%%%%%%%%%%%%%%%%%%%%%%%%%%%%%%%%%%%%%%%%%%%%%%%%%%%%%%%%%%%%
\section*{Study of photon conversions}

Photon conversions occurring in the material inside the TEC may 
cause additional tracks and are studied on data and Monte Carlo for 
each year independently.
%The identification of photon conversions 
%is done using data and Monte Carlo samples for each year. 
A loose selection of \taupg~ events is made, requiring two
low multiplicity jets and the cosine of the polar angle of the event
thrust axis $|\cos \theta_{thrust}| <$ 0.7.
Radiative photons or photons from $\pi^0$ decays can convert in the
detector. The tracks from the conversion point either to their corresponding cluster
in the BGO calorimeter, or to a coalescent cluster including the energy of the two
conversion tracks or of the 2 photons in the case of a $\pi^0$ decay.
Therefore, the $p_T$ measured in the central tracker must be smaller than
the transverse energy observed in the electromagnetic calorimeter.

\begin{figure}[htbp]
\begin{center}
\mbox {\epsfig{figure=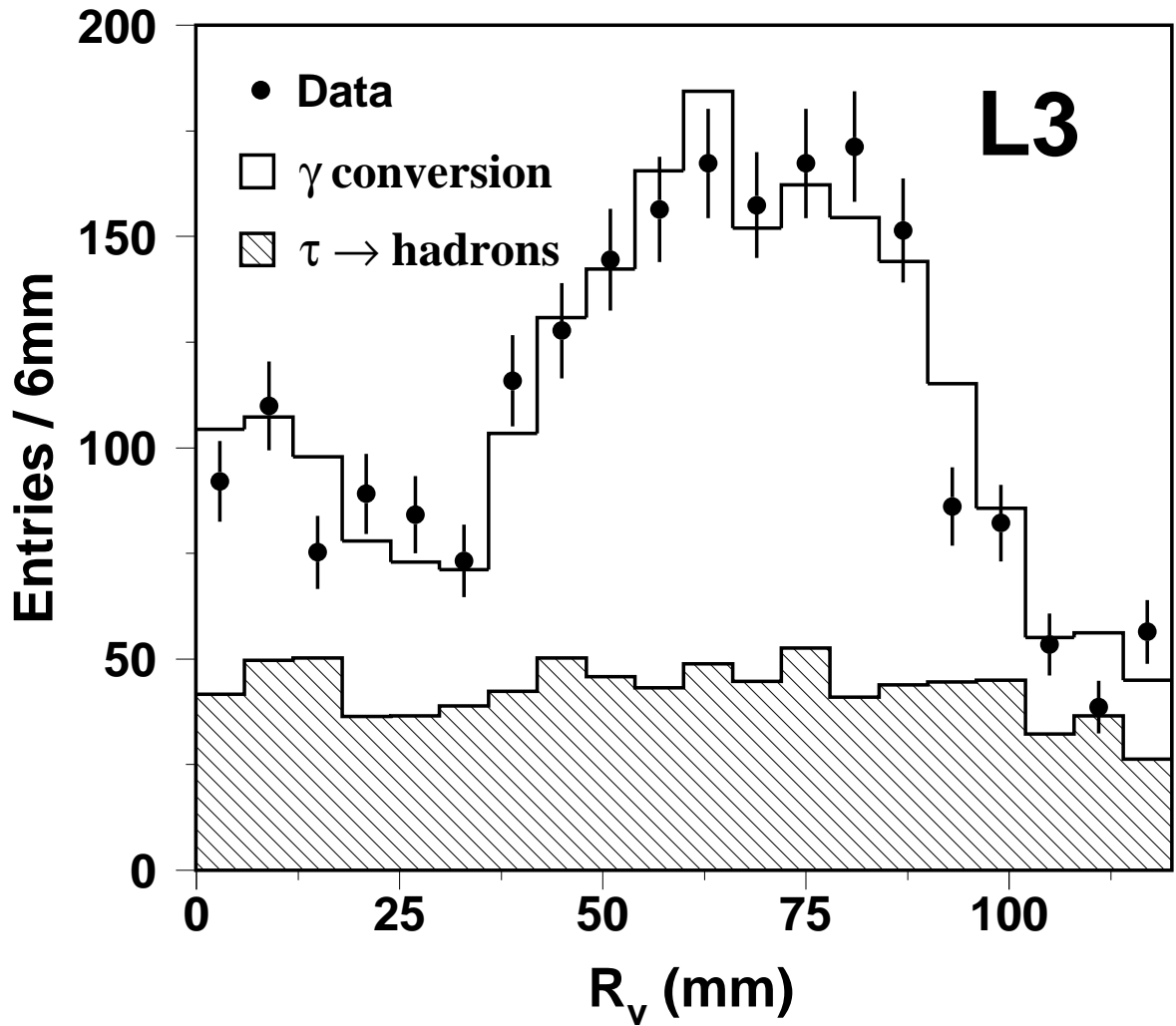,width=0.92\textwidth}}
\end{center}
\caption[]{\label{fig:gconv} Distribution of the radial distance
from the beam axis, $R_v$, of vertices
reconstructed using photon conversion tracks.
The flat hatched distribution
stems from track pairs of hadronic $\tau$ decays.}
\end{figure}

In the case that only one track reaches the TEC, its 
transverse momentum must be
less than 4$\GeV$. When both tracks are
reconstructed, 
the square of their invariant mass
must be less than 0.005 \GeV$^2$.
Taking track pairs which fulfil these requirements, the distance of 
their vertex to the beam axis, $R_v$, is calculated. The distribution
of $R_v$ is shown in Figure~\ref{fig:gconv}
for data from 1994 and Monte Carlo:
most of the photon conversions occur at radii between 40 and 90 mm,
corresponding to the
position of the two 
cylindrical layers of the SMD.
Good agreement of the simulation of 
the photon conversion probability inside the TEC with the data is obtained
after enlarging the conversion probability by a factor of about 1.6 for 
data taking periods after the installation of the SMD, to 
account for additional material not fully considered in the Monte Carlo simulation. 
The flat background stems mainly from 3-prong hadronic $\tau$ decays.

After rejection of identified photon conversions,
$ 0.4\%$
of the $\tau $
decays still contain tracks from photon conversions. They are accounted
for in the migration efficiencies determined from Monte Carlo. 

The effect of tracks scattered back from BGO 
clusters is investigated.
As their momenta are low
%and they do not point to the interaction
%region and 
they are removed by the requirement on the
transverse momentum of a track.

%%%%%%%%%%%%%%%%%%%%%%%%%%%%%%%%%%%%%%%%%%%%%%%%%%%%%%%%%%%%%%%%%%%%%%%%%%%%%%%
% Events selection
%%%%%%%%%%%%%%%%%%%%%%%%%%%%%%%%%%%%%%%%%%%%%%%%%%%%%%%%%%%%%%%%%%%%%%%%%%%%%%%
\section*{Selection of \boldtaupg~events }

Events of the process \taupg~are characterised by
two jets with low track and calorimetric cluster multiplicities,
where a jet may consist of an
isolated electron or muon.
To ensure good track measurements only events in the 
barrel region 
of the detector are accepted by requiring
$|\cos \theta_{thrust}| <$ 0.7.
The event multiplicity is defined as the sum of the number of tracks in the TEC 
and the number of neutral calorimetric clusters, without an assigned charged track
and with an energy larger than $0.5 \GeV$. 
This event multiplicity is required to be less than 10.
Each event is divided into two hemispheres with respect to the plane orthogonal
to the thrust axis. 
The main backgrounds arise from two-photon interactions, $\rm e^+e^-$ events and Z decays into
two muons. These processes are rejected using information from both hemispheres.
\begin{itemize}

\item 
Two-photon interactions:
each hemisphere must contain at least one calorimetric 
cluster.
The sum of the energy
deposited in the calorimeters and the muon momenta
measured in the muon spectrometer must be larger
than 13$\GeV$.

\item 
$\rm e^+e^-$ events: the total energy deposited in the BGO must
be less than
60$\GeV$. In addition,  the energy deposit in each hemisphere must be 
less than 44$\GeV$
and  the acoplanarity angle between the leading tracks
of the two hemispheres must be larger than
0.003 rad. 

\item Z decays into two muons:
events with a track in the muon chambers
with a momentum larger than 
42 $\GeV$ are not accepted.
Furthermore, events with a muon or a MIP in both hemispheres 
are rejected.

\end{itemize}

A sample of 70016 \taupg~
events is selected.
The estimations of the efficiencies and background fractions are done 
separately for each year.
The average selection efficiency for $\taupg$ events
inside the barrel,
estimated from Monte
Carlo, is
78.8 $\pm$ 0.2\%.

\begin{figure}[hp]
$\begin{array}{cc}
\mbox {\epsfig{figure=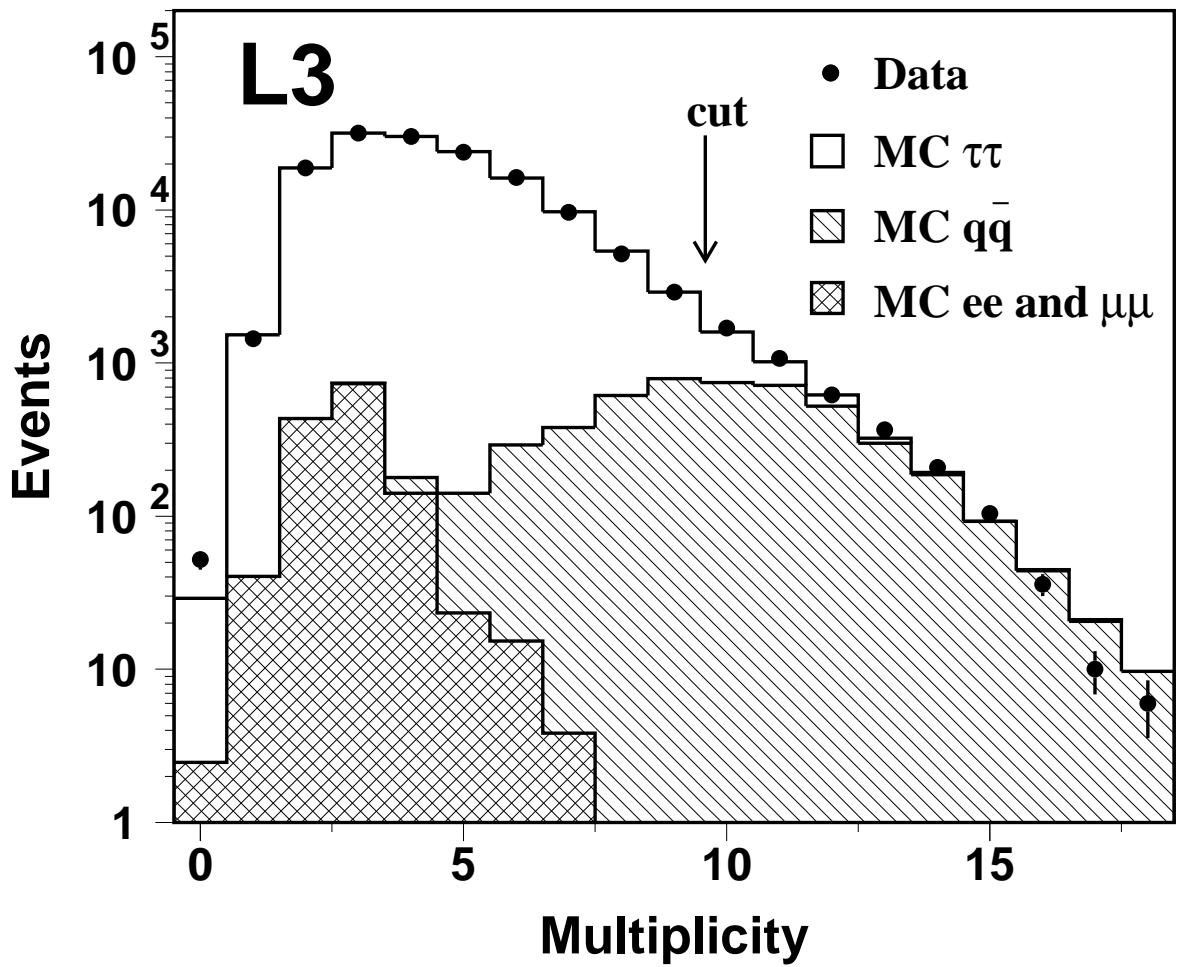,width=0.92\textwidth}}
\end{array}$
\caption[]{\label{fig:nb05end} The
distribution of the event multiplicity.
The Monte Carlo prediction
for $\taupg$
and the background from other leptonic and hadronic Z decays
after application of the scale factors
is also given.
}
\end{figure}

\begin{figure}[hp]
$\begin{array}{cc}
\mbox {\epsfig{figure=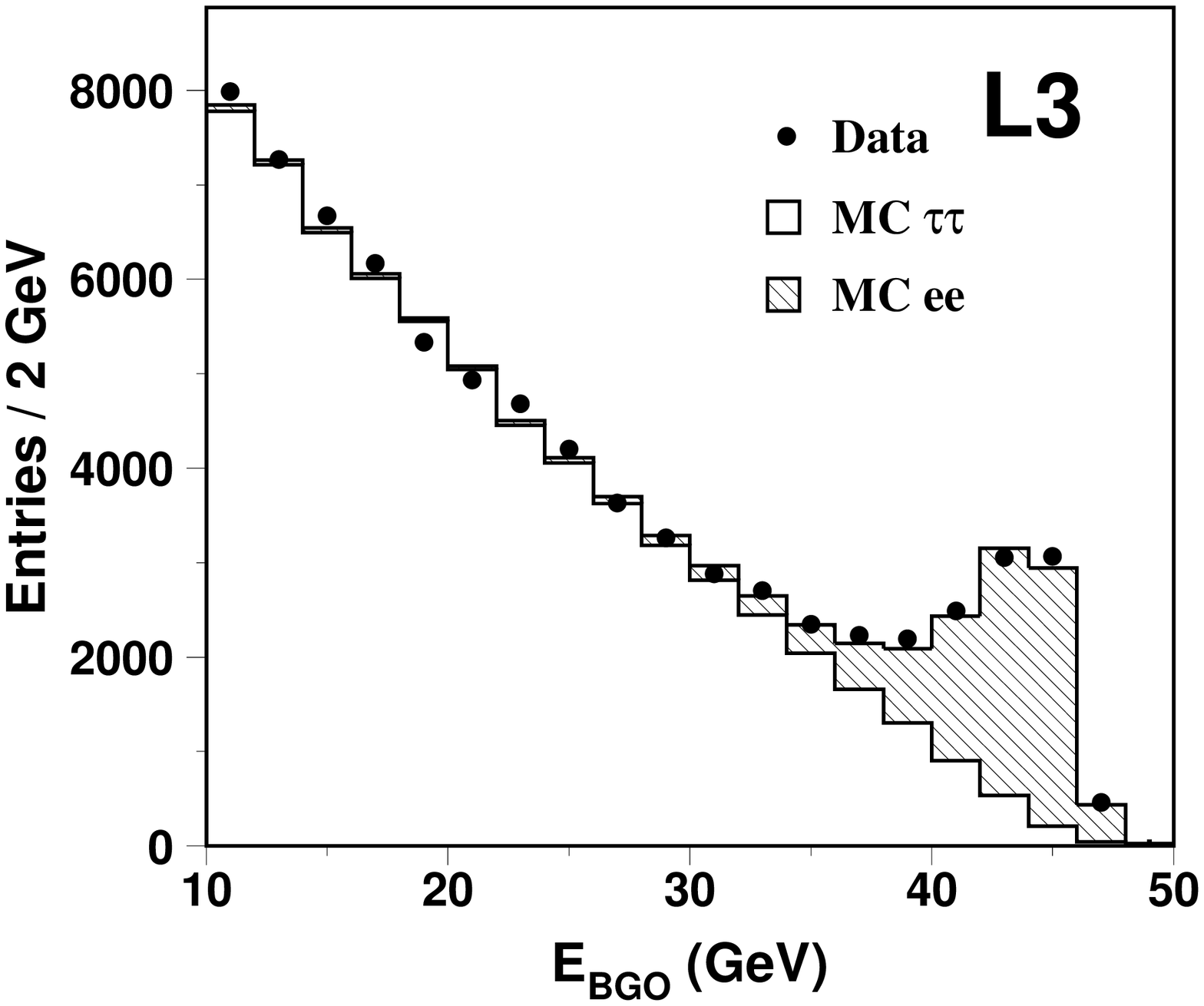,width=0.92\textwidth}}
\end{array}$
\caption[]{\label{fig:ebend} The distribution of the BGO energy
for selected  $\taupg$
events with relaxed cuts against $\eee$ background.
Also shown is the Monte Carlo expectation 
for $\taupg$ 
and $\eee$ events after rescaling.}
\end{figure}

\begin{figure}[hp]
$\begin{array}{cc}
\mbox {\epsfig{figure=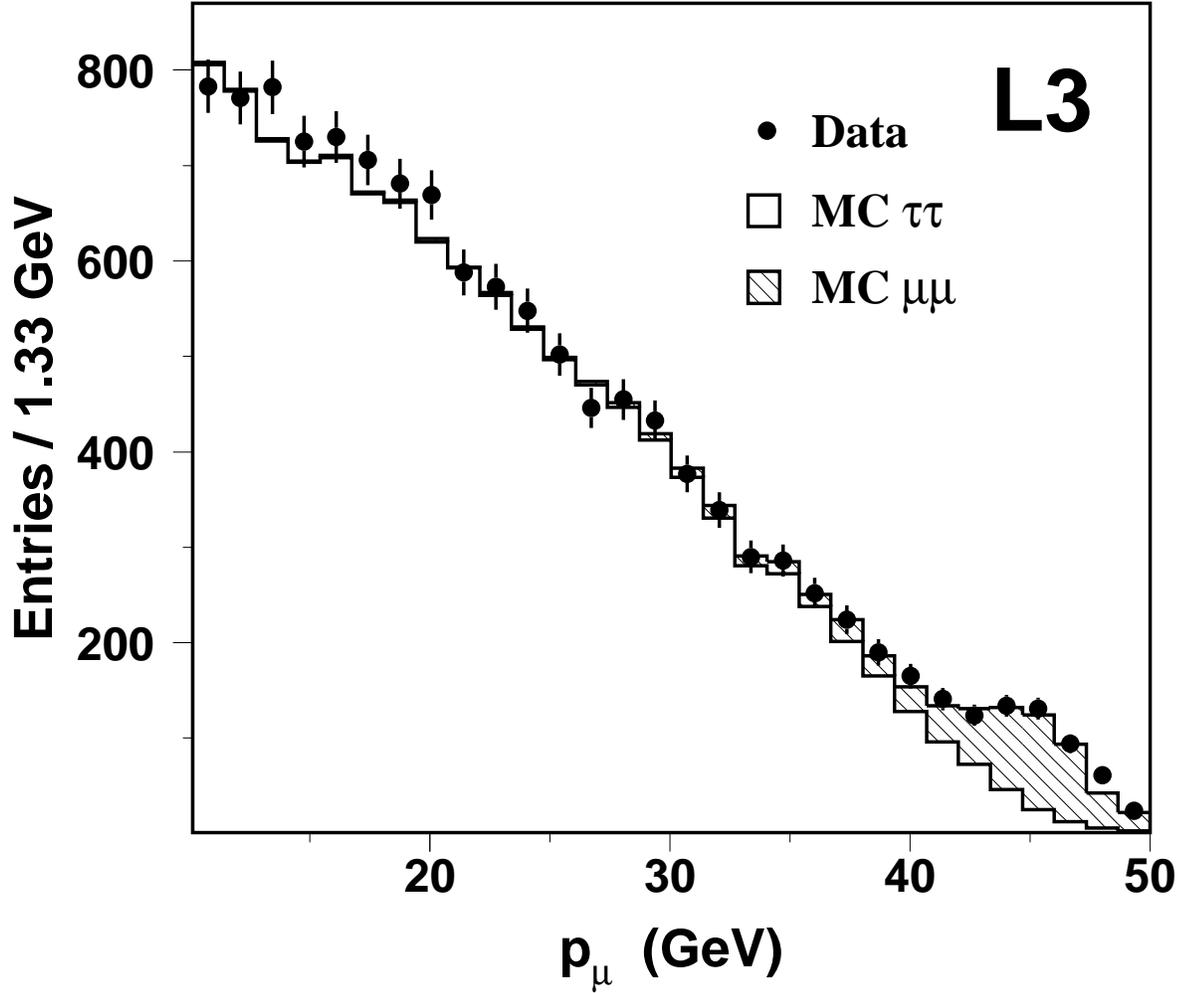,width=0.92\textwidth}}
\end{array}$
\caption[]{\label{fig:puend} The momentum distribution of tracks
measured in the muon chambers in the $\taupg$ event sample
with relaxed cuts against dimuon background.
Also shown is the Monte Carlo expectation 
for $\taupg$
and the background from dimuon final states after rescaling. 
}
\end{figure}

The background from hadronic and other leptonic Z decays
and two-photon interactions
is estimated from Monte Carlo.
%and amounts to  1.94\% and $0.15\%$,
%respectively.
As an example, the distribution of the event multiplicity,
is shown in Figure~\ref{fig:nb05end} for data 
and a superposition of Monte Carlo
from \taupg~
and background. The background from
hadronic Z decays dominates at large values of the multiplicity.
After applying a correction factor to 
the fraction of the hadronic background of 1.05,
very good agreement is obtained. 
The background from Bhabha events is determined 
using the upper part of the energy distribution measured in the BGO.
This is shown in Figure~\ref{fig:ebend} 
for an event sample with the Bhabha rejection cuts relaxed.
The prediction from Bhabha Monte Carlo is scaled
by a factor of 1.1 to agree with the data. 
The background after the final selection
is estimated from Bhabha events using this scale factor.
% to be
%$0.16$\%. 
The same procedure is applied to estimate the 
$\mu^+\mu^-$ background.
% which is found to be  $0.68\%$. 
Figure~\ref{fig:puend}
shows the spectrum of muons measured in the muon spectrometer
for data and Monte Carlo after applying a correction factor of 1.1.
The background fractions from all sources are
listed in Table~\ref{table:backg}. 
The cosmic background,
estimated from the distribution 
of the distance of closest approach
to the beam position, is found to be negligible.

%%%%%%%%%%%%%%%%%%%%%%%%%%%%%%%%%%%%%%%%%%%%%%%%%%%%%%%%%%%%%%%%%%%%%%%%%%%%%%%
% Determination of the topological branching fractions
%%%%%%%%%%%%%%%%%%%%%%%%%%%%%%%%%%%%%%%%%%%%%%%%%%%%%%%%%%%%%%%%%%%%%%%%%%%%%%%
\section*{Determination of the topological branching fractions}

The maximum likelihood method is used 
to determine the topological branching fractions,
with likelihood function $L$ defined as:
\begin{eqnarray}
  L =  \prod_{i=0}^6 P(N^i_{obs},~N^i_{exp}),
\end{eqnarray}
\noindent
where $P$ is the Poisson distribution, $N_{obs}^i$ is the number of observed
events and $N_{exp}^i$ is the number of expected events
with $i$ reconstructed tracks. 
The latter is:
\begin{eqnarray}
N^i_{exp} = N_{\tau} \sum_{j=1,3,5} \br (j) \varepsilon^{ij} 
+ \sum_k N_{bg}^{ik},
\end{eqnarray}
\noindent
where
$N_{\tau}$ is the number of $\tau$ decays
and $\br(j)$ is the branching fraction of
$j$-prong $\tau$ decays, $j$ = 1, 3 or 5.
The elements of the track detection efficiency matrix, $\varepsilon^{ij}$, 
are determined from Monte Carlo. The non-diagonal elements
represent migrations
between the topologies.
The number of non-tau background events, $N_{bg}^{ik}$, 
obtained from Monte Carlo, is
normalised to the data
luminosity. The index $k$ runs over all background sources.

The efficiency matrix of the track reconstruction is shown in 
Table~\ref{table:eff}, indicating the numbers of reconstructed tracks 
for $\tau$ decays to 1, 3 and 5 charged particles. 
Tracks arising from neutral kaon decays are not accounted for in the topology.
Table~\ref{table:obs} shows the number of observed 
$\tau$ decays in the different
topologies and the estimated background. 
In the fit 
the constraint $\br(1)+\br(3)+\br(5)=1$
%$\brts{3} = 1 - \brts{1}
%-\brts{5} $ 
is applied
and the sum of $N^i_{exp}$ is constrained to the number of
observed $\tau$ decays.
The following results 
are obtained:
\begin{eqnarray*}
 & \brts{1} &  
                      = 85.274 \pm 0.105 \%, \\
 & \brts{3} &
                      =  14.556 \pm 0.105 \%, \\  
 & \brts{5} &
                      = \pho 0.170 \pm 0.022 \%,  
\end{eqnarray*}
where the uncertainty is statistical only. 
The $\chi^2$/d.o.f. is 5.7/4.
The correlation coefficients
are given in Table~\ref{table:cor}.

\begin{figure}[hp]
$\begin{array}{cc}
\mbox {\epsfig{figure=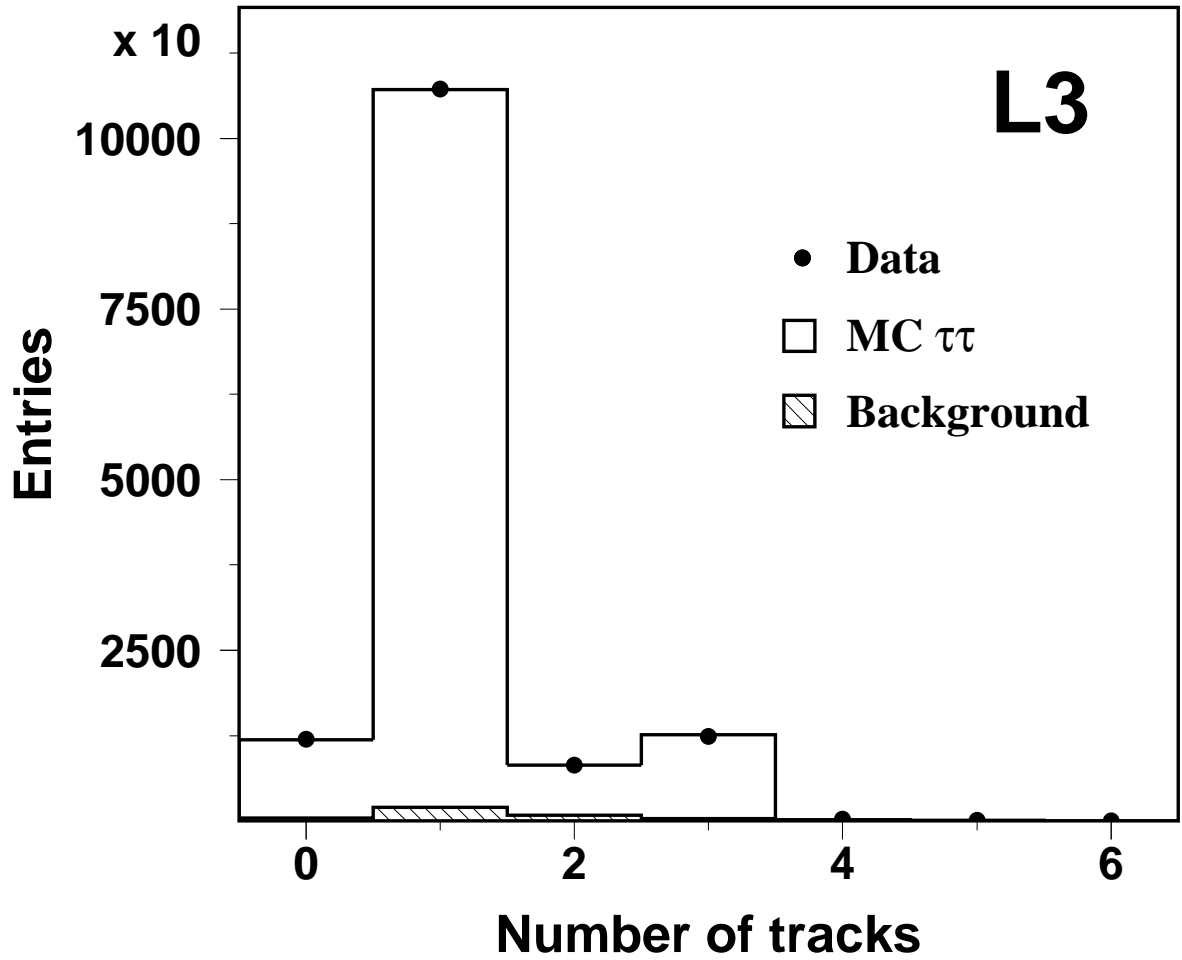,width=0.92\textwidth}}
\end{array}$
\caption[]{\label{fig:trklin} The charged multiplicity distribution from
$\tau$ decays. Also shown is the expectation 
from Monte Carlo for $\taupg$
 and the background.
  }
\end{figure}

\begin{figure}[hp]
$\begin{array}{cc}
\mbox {\epsfig{figure=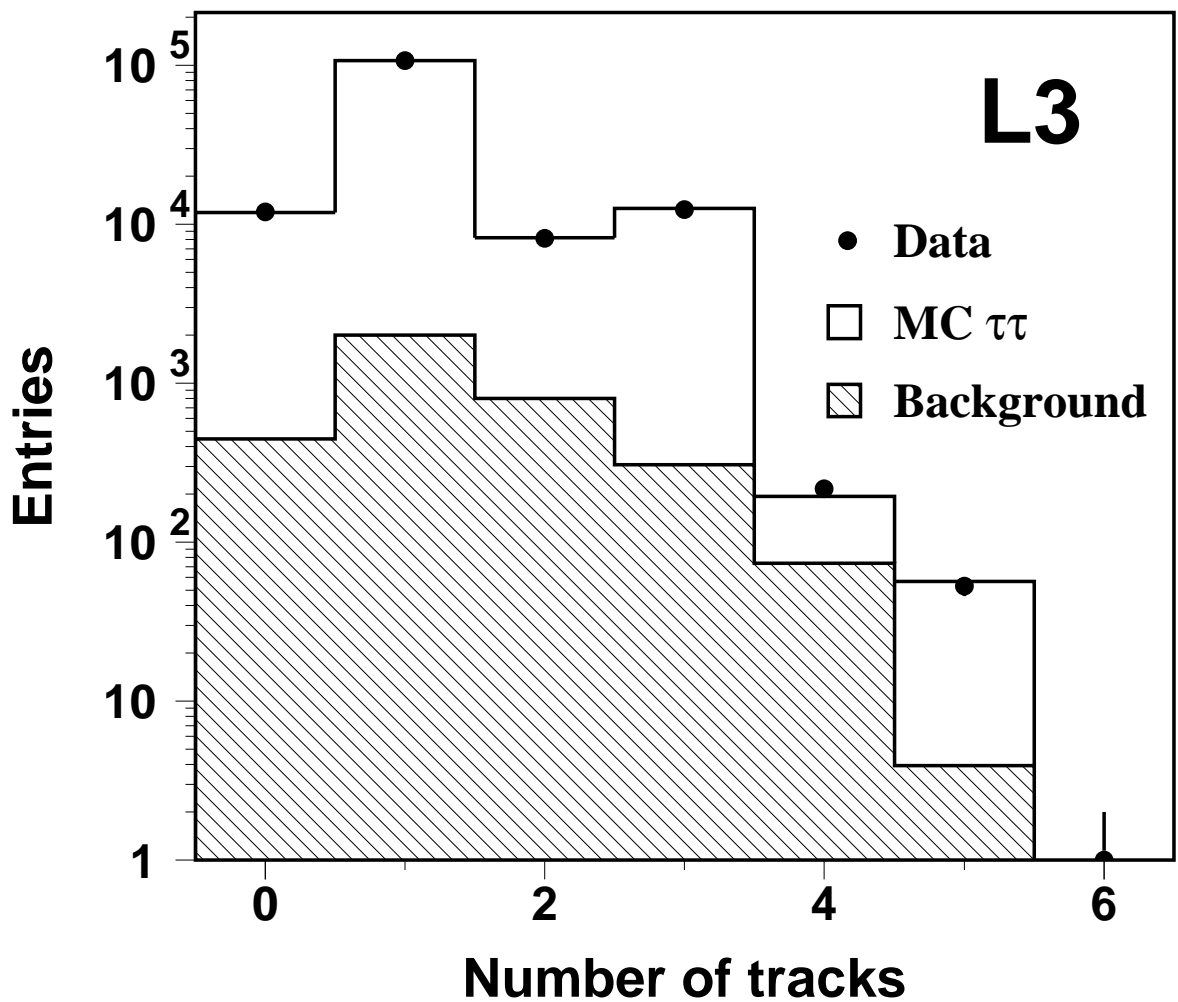,width=0.92\textwidth}}
\end{array}$
\caption[]{\label{fig:trklog} The charged multiplicity distribution from
$\tau$ decays.
Also shown is the expectation 
from Monte Carlo for $\taupg$ and the background
from hadronic Z decays and other sources.
  }
\end{figure}

In Figures~\ref{fig:trklin} and~\ref{fig:trklog}
the number of observed $\tau$ decays is shown
as a function of the charged track multiplicity
in linear and logarithmic scale, respectively.
Also shown are the Monte Carlo predictions,
using the fitted branching fractions, and the 
background.

%%%%%%%%%%%%%%%%%%%%%%%%%%%%%%%%%%%%%%%%%%%%%%%%%%%%%%%%%%%%%%%%%%%%%%%%%%%%%%%
% Systematic
%%%%%%%%%%%%%%%%%%%%%%%%%%%%%%%%%%%%%%%%%%%%%%%%%%%%%%%%%%%%%%%%%%%%%%%%%%%%%%%
\section*{Systematic uncertainties}

The criteria to suppress the different background sources
are varied within reasonable ranges and the 
changes in the topological branching fractions
are taken as systematic uncertainties.
The uncertainty on the cross section of 
e$^+$e$^-\rightarrow$ hadrons
\cite{l3_zres}
and the scale factor applied to the Monte Carlo
normalisation are considered.
The cross section 
uncertainties on $\rm e^+e^-$ and $\mu^+\mu^-$ final states and two-photon interactions
have negligible effects on the branching fractions.
The background uncertainties from $\rm e^+e^-$ and $\mu^+\mu^-$ final states 
are obtained from the statistical uncertainty of 
1\% on the scale factors
applied to the Monte Carlo distributions shown in Figures
~\ref{fig:ebend} and~\ref{fig:puend}. 
The systematic uncertainty due to track efficiency is obtained by
varying this quantity within
its statistical uncertainty of 0.25\%.
Furthermore, the track definition criteria 
%and the size of the anode region 
are changed within
reasonable ranges. 
The uncertainty from double track resolution is estimated
by reweighting the $\Delta \phi$ distribution of Figure~\ref{fig:dtr}
forcing agreement between data and Monte Carlo  
in the low $\Delta \phi$ region. The effects on the branching fractions are
taken as systematic uncertainties.

The systematic uncertainty due to photon conversions is
obtained from the statistical uncertainty of 10\% on the
photon conversion probability correction factor and
from variations of conversion identification criteria.

The uncertainty from Monte Carlo statistics is included as
statistical uncertainties on the efficiency matrix in Table 1. The
resulting variations on the branching fractions are taken as systematic
uncertainties.
Effects from the energy scale uncertainties
of the subdetectors are negligible. 
A summary of the systematic uncertainties 
is provided in Table~\ref{table:sys}. 
%%%%%%%%%%%%%%%%%%%%%%%%%%%%%%%%%%%%%%%%%%%%%%%%%%%%%%%%%%%%%%%%%%%%%%%%%%%%%%%
% Summary
%%%%%%%%%%%%%%%%%%%%%%%%%%%%%%%%%%%%%%%%%%%%%%%%%%%%%%%%%%%%%%%%%%%%%%%%%%%%%%%

After combination of the systematic uncertainties
the results for
the branching fractions of the 
$\tau$ lepton decays into one, three and five
charged particle final states are:
\begin{eqnarray*}
 & \brt{1} &  
                      = 85.274 \pm 0.105 \pm 0.073 \% , \\
 & \brt{3} &
                      =  14.556 \pm 0.105 \pm 0.076 \%, \\  
 & \brt{5} &
                      = \pho 0.170 \pm 0.022 \pm 0.026 \%,
\end{eqnarray*}
where the first uncertainty is statistical
and the second is systematic.
These new results are in agreement with a recent measurement with 
the full LEP statistics~\cite{delphip} and with the 
current world averages~\cite{PDG2000}.

%%%%%%%%%%%%%%%%%%%%%%%%%%%%%%%%%%%%%%%%%%%%%%%%%%%%%%%%%%%%%%%%%%%%%%%%%%%%%%%

\section*{Acknowledgements}

We wish to express our  gratitude to the CERN  accelerator  division for
the  excellent  performance  of the  LEP  machine.  We  acknowledge  the
contributions of engineers and technicians who have participated
in the construction and maintenance of this experiment.

%%%%%%%%%%%%%%%%%%%%%%%%%%%%%%%%%%%%%%%%%%%%%%%%%%%%%%%%%%%%%%%%%%%%%%%%%%%%%%%
% Bibliography
%%%%%%%%%%%%%%%%%%%%%%%%%%%%%%%%%%%%%%%%%%%%%%%%%%%%%%%%%%%%%%%%%%%%%%%%%%%%%%
% Style file to use with mcite.
% Use l3style with just cite.

\newpage

\bibliographystyle{l3stylem}

\newpage
%%%%%%%%%%%%%%%%%%%%%%%%%%%%%%%%%%%%%%%%%%%%%%%%%%%%%%%%%%%%%%%%%%%%%%%%%%%%%%%
% The author list
%%%%%%%%%%%%%%%%%%%%%%%%%%%%%%%%%%%%%%%%%%%%%%%%%%%%%%%%%%%%%%%%%%%%%%%%%%%%%%%
%
%\section*{Author List}
%
\newpage
\typeout{   }     
\typeout{Use only for paper240 - tau paper? }
\typeout{Use only for paper240 - tau paper? }
\typeout{Use only for paper240 - tau paper? }
\typeout{Use only for paper240 - tau paper? }
\typeout{Use only for paper240 - tau paper? }
\typeout{Use only for paper240 - tau paper? }
\typeout{Use only for paper240 - tau paper? }
\typeout{Use only for paper240 - tau paper? }
\typeout{Use only for paper240 - tau paper? }
\typeout{Use only for paper240 - tau paper? }
\typeout{$Modified: Jul 3 2001 by smele $}
\typeout{!!!!  This should only be used with document option a4p!!!!}
\typeout{   }
%
%
%
%  L A T E X  version!!
%
%
% Make sure that the Lep package has been used!
%\input{Lep.sty}%
%
%\ifx\LepCalled\undefined%
%\typeout{     }%
%\typeout{!!!!!!!!!!!!!!!!!!!!!!!!!!!!!!!!!!!!!!!!!!!!!!!!!!!!!!!!!!!}%
%\typeout{Yikes.  You haven't used the Lep package!}%
%\typeout{Please put \protect\usepackage\protect{Lep\protect} in your preamble,
%         followed by}%
%\typeout{\protect\Lep\protect{1\protect} or \protect\Lep\protect{2\protect}}%
%\typeout{     }%
%\typeout{For now you will get a Lep phase 2 authorlist (may not be right!).}%
%\typeout{!!!!!!!!!!!!!!!!!!!!!!!!!!!!!!!!!!!!!!!!!!!!!!!!!!!!!!!!!!!}%
%\typeout{     }%
%\Lep{2}\fi%

\newcount\tutecount  \tutecount=0
\def\tutenum#1{\global\advance\tutecount by 1 \xdef#1{\the\tutecount}}
\def\tute#1{$^{#1}$}
\tutenum\aachen            % 1
\tutenum\nikhef            % 2
\tutenum\mich              % 3
\tutenum\lapp              % 4
\tutenum\basel             % 5
\tutenum\lsu               % 6
\tutenum\beijing           % 7
\tutenum\berlin            % 8
\tutenum\bologna           % 9 
\tutenum\tata              % 10
\tutenum\ne                % 11
\tutenum\bucharest         % 12
\tutenum\budapest          % 13
\tutenum\mit               % 14 
\tutenum\panjab            % 15
\tutenum\debrecen          % 16
\tutenum\florence          % 17
\tutenum\cern              % 18
\tutenum\wl                % 19
\tutenum\geneva            % 20
\tutenum\hefei             % 21
\tutenum\lausanne          % 22
\tutenum\lyon              % 23
\tutenum\madrid            % 24
\tutenum\florida           % 25
\tutenum\milan             % 26
\tutenum\moscow            % 27
\tutenum\naples            % 29
\tutenum\cyprus            % 30
\tutenum\nymegen           % 31
\tutenum\caltech           % 32
\tutenum\perugia           % 33
\tutenum\peters            % 34
\tutenum\cmu               % 35
\tutenum\potenza           % 36
\tutenum\prince            % 37
\tutenum\riverside         % 38
\tutenum\rome              % 39
\tutenum\salerno           % 40
\tutenum\ucsd              % 41
\tutenum\sofia             % 42
\tutenum\korea             % 43
\tutenum\utrecht           % 44
\tutenum\purdue            % 45
\tutenum\psinst            % 46
\tutenum\zeuthen           % 47
\tutenum\eth               % 48
\tutenum\hamburg           % 49
\tutenum\taiwan            % 50
\tutenum\tsinghua          % 51

{
\parskip=0pt
\noindent
{\bf The L3 Collaboration:}
\ifx\selectfont\undefined%  old style font selection
 \baselineskip=10.8pt
 \baselineskip\baselinestretch\baselineskip
 \normalbaselineskip\baselineskip
 \ixpt
\else%                      new style font selection
 \fontsize{9}{10.8pt}\selectfont
\fi
\medskip
\tolerance=10000
\hbadness=5000
\raggedright
\hsize=162truemm\hoffset=0mm
\def\r{\rlap,}
\noindent

P.Achard\r\tute\geneva\ 
O.Adriani\r\tute{\florence}\ 
M.Aguilar-Benitez\r\tute\madrid\ 
J.Alcaraz\r\tute{\madrid,\cern}\ 
G.Alemanni\r\tute\lausanne\
J.Allaby\r\tute\cern\
A.Aloisio\r\tute\naples\ 
M.G.Alviggi\r\tute\naples\
H.Anderhub\r\tute\eth\ 
V.P.Andreev\r\tute{\lsu,\peters}\
% ...
\newpage
%\rule{\textwidth}{0.4pt}
\begin{list}{A}{\itemsep=0pt plus 0pt minus 0pt\parsep=0pt plus 0pt minus 0pt
                \topsep=0pt plus 0pt minus 0pt}
\item[\aachen]
 I. Physikalisches Institut, RWTH, D-52056 Aachen, FRG$^{\S}$\\
 III. Physikalisches Institut, RWTH, D-52056 Aachen, FRG$^{\S}$
\item[\nikhef] National Institute for High Energy Physics, NIKHEF, 
     and University of Amsterdam, NL-1009 DB Amsterdam, The Netherlands
\item[\mich] University of Michigan, Ann Arbor, MI 48109, USA
\item[\lapp] Laboratoire d'Annecy-le-Vieux de Physique des Particules, 
     LAPP,IN2P3-CNRS, BP 110, F-74941 Annecy-le-Vieux CEDEX, France
\item[\basel] Institute of Physics, University of Basel, CH-4056 Basel,
     Switzerland
\item[\lsu] Louisiana State University, Baton Rouge, LA 70803, USA
\item[\beijing] Institute of High Energy Physics, IHEP, 
  100039 Beijing, China$^{\triangle}$ 
\item[\berlin] Humboldt University, D-10099 Berlin, FRG$^{\S}$
\item[\bologna] University of Bologna and INFN-Sezione di Bologna, 
     I-40126 Bologna, Italy
\item[\tata] Tata Institute of Fundamental Research, Mumbai (Bombay) 400 005, India
\item[\ne] Northeastern University, Boston, MA 02115, USA
\item[\bucharest] Institute of Atomic Physics and University of Bucharest,
     R-76900 Bucharest, Romania
\item[\budapest] Central Research Institute for Physics of the 
     Hungarian Academy of Sciences, H-1525 Budapest 114, Hungary$^{\ddag}$
\item[\mit] Massachusetts Institute of Technology, Cambridge, MA 02139, USA
\item[\panjab] Panjab University, Chandigarh 160 014, India.
\item[\debrecen] KLTE-ATOMKI, H-4010 Debrecen, Hungary$^\P$
\item[\florence] INFN Sezione di Firenze and University of Florence, 
     I-50125 Florence, Italy
\item[\cern] European Laboratory for Particle Physics, CERN, 
     CH-1211 Geneva 23, Switzerland
\item[\wl] World Laboratory, FBLJA  Project, CH-1211 Geneva 23, Switzerland
\item[\geneva] University of Geneva, CH-1211 Geneva 4, Switzerland
\item[\hefei] Chinese University of Science and Technology, USTC,
      Hefei, Anhui 230 029, China$^{\triangle}$
\item[\lausanne] University of Lausanne, CH-1015 Lausanne, Switzerland
\item[\lyon] Institut de Physique Nucl\'eaire de Lyon, 
     IN2P3-CNRS,Universit\'e Claude Bernard, 
     F-69622 Villeurbanne, France
\item[\madrid] Centro de Investigaciones Energ{\'e}ticas, 
     Medioambientales y Tecnolog{\'\i}cas, CIEMAT, E-28040 Madrid,
     Spain${\flat}$ 
\item[\florida] Florida Institute of Technology, Melbourne, FL 32901, USA
\item[\milan] INFN-Sezione di Milano, I-20133 Milan, Italy
\item[\moscow] Institute of Theoretical and Experimental Physics, ITEP, 
     Moscow, Russia
\item[\naples] INFN-Sezione di Napoli and University of Naples, 
     I-80125 Naples, Italy
\item[\cyprus] Department of Physics, University of Cyprus,
     Nicosia, Cyprus
\item[\nymegen] University of Nijmegen and NIKHEF, 
     NL-6525 ED Nijmegen, The Netherlands
\item[\caltech] California Institute of Technology, Pasadena, CA 91125, USA
\item[\perugia] INFN-Sezione di Perugia and Universit\`a Degli 
     Studi di Perugia, I-06100 Perugia, Italy   
\item[\peters] Nuclear Physics Institute, St. Petersburg, Russia
\item[\cmu] Carnegie Mellon University, Pittsburgh, PA 15213, USA
\item[\potenza] INFN-Sezione di Napoli and University of Potenza, 
     I-85100 Potenza, Italy
\item[\prince] Princeton University, Princeton, NJ 08544, USA
\item[\riverside] University of Californa, Riverside, CA 92521, USA
\item[\rome] INFN-Sezione di Roma and University of Rome, ``La Sapienza",
     I-00185 Rome, Italy
\item[\salerno] University and INFN, Salerno, I-84100 Salerno, Italy
\item[\ucsd] University of California, San Diego, CA 92093, USA
\item[\sofia] Bulgarian Academy of Sciences, Central Lab.~of 
     Mechatronics and Instrumentation, BU-1113 Sofia, Bulgaria
\item[\korea]  The Center for High Energy Physics, 
     Kyungpook National University, 702-701 Taegu, Republic of Korea
\item[\utrecht] Utrecht University and NIKHEF, NL-3584 CB Utrecht, 
     The Netherlands
\item[\purdue] Purdue University, West Lafayette, IN 47907, USA
\item[\psinst] Paul Scherrer Institut, PSI, CH-5232 Villigen, Switzerland
\item[\zeuthen] DESY, D-15738 Zeuthen, 
     FRG
\item[\eth] Eidgen\"ossische Technische Hochschule, ETH Z\"urich,
     CH-8093 Z\"urich, Switzerland
\item[\hamburg] University of Hamburg, D-22761 Hamburg, FRG
\item[\taiwan] National Central University, Chung-Li, Taiwan, China
\item[\tsinghua] Department of Physics, National Tsing Hua University,
      Taiwan, China
\item[\S]  Supported by the German Bundesministerium 
        f\"ur Bildung, Wissenschaft, Forschung und Technologie
\item[\ddag] Supported by the Hungarian OTKA fund under contract
numbers T019181, F023259 and T024011.
\item[\P] Also supported by the Hungarian OTKA fund under contract
  number T026178.
\item[$\flat$] Supported also by the Comisi\'on Interministerial de Ciencia y 
        Tecnolog{\'\i}a.
\item[$\sharp$] Also supported by CONICET and Universidad Nacional de La Plata,
        CC 67, 1900 La Plata, Argentina.
\item[$\natural$] Also supported by Deutscher akademischer 
Austauschdienst.
\item[$\triangle$] Supported by the National Natural Science
  Foundation of China.
\end{list}
}
\vfill

%%% Local Variables: 
%%% mode: latex
%%% TeX-master: t
%%% End:

%%%%%%%%%%%%%%%%%%%%%%%%%%%%%%%%%%%%%%%%%%%%%%%%%%%%%%%%%%%%%%%%%%%%%%%%%%%%%%%
%\subsection*{Figures}
%%%%%%%%%%%%%%%%%%%%%%%%%%%%%%%%%%%%%%%%%%%%%%%%%%%%%%%%%%%%%%%%%%%%%%%%%%%%%%%

\begin{table}
\begin{center}
\begin{tabular}{|c|c|} \hline
Background source      & Fraction [\%] \\ \hline
$ {\rm Z}\rightarrow$ hadrons &  1.59        \\
Two-photon interactions   &  0.16        \\  
\eee                   &  0.16        \\
\emm                   &  0.68        \\ 
\hline
\end{tabular}
\caption{The background fractions from the different sources
in the \taupg~ event sample.  
}
\label{table:backg}
\end{center}
\end{table}

\begin{table}
\begin{center}
\begin{tabular}{|c|c|c|c|} \cline{2-4}
\multicolumn{1}{c}{}&\multicolumn{3}{|c|}{$N_{gen}$}    \\  \hline 
 $N_{rec}$  &  1  &  3  &  5       \\  \hline
   0       &\pho $7.58\pm 0.02$& \pho $0.78\pm 0.01$& \pho $0.84\pm 0.15$ \\       
   1       & $70.18\pm 0.05$& \pho $6.88\pm 0.04$&    \pho $3.58\pm 0.30$ \\ 
   2       &\pho $0.33\pm 0.01$&$26.89\pm 0.07$&      \pho $9.55\pm 0.50$ \\
   3       &\pho $0.16\pm 0.01$&$47.05\pm 0.09$&$20.50\pm 0.73$\\
   4       & $ <$0.01     & \pho $0.21\pm 0.01$&$24.38\pm 0.79$ \\
   5       & $ <$0.01     & \pho $0.05\pm 0.01$&$14.88\pm 0.63$ \\
   6       & $ <$0.01     & $ <$0.01     &    $ <$0.01          \\
\hline
\end{tabular}
\caption{The efficiency matrix of track reconstruction in percent.
$N_{gen}$ denotes the number of charged tracks
of the $\tau$ decay before detector simulation and $N_{rec}$  
the number of tracks after the reconstruction.}
\label{table:eff}
\end{center}
\end{table}

\begin{table}
\begin{center}
\begin{tabular}{|c|ccccccc|} \hline
$N_{rec}$  & 0 & 1  & 2 & 3 & 4& 5 & 6 \\ \hline
Data & 11935& 107283&8166&12378&216&53&1\\ \hline
Background [\%]& 3.7 &  1.9 &  9.7 & 2.5 & 33.8& 7.6 & 0 \\  
\hline
\end{tabular}
\caption{ The number of $\tau$ decays observed in the different topologies  
and background estimated by Monte Carlo.}
\label{table:obs}
\end{center}
\end{table}

\begin{table}
\begin{center}
\begin{tabular}{|l|c|c|c|} \hline
Source     & $\brts{1}$  & $\brts{3}$ &  $\brts{5}$ \\ \hline
$\brts{1}$ & $\phantom{-}1.0\phantom{00}$  & $-0.978$ & $-0.082$   \\ \hline
$\brts{3}$ &       & $\phantom{-}1.0\phantom{00}$       & $-0.127$ \\  \hline
$\brts{5}$ &       &             &  $\phantom{-}1.0\phantom{00}$     \\ \hline
\end{tabular}
\caption{The correlation coefficients obtained from a fit
of the topological branching fractions.}
\label{table:cor}
\end{center}
\end{table}

%\begin{table}[ht]
\begin{table}
\begin{center}
\begin{tabular}{|l|c|c|c|} \hline
Source                        & $\brts{1}$     &  $\brts{3}$    & $\brts{5}$     \\ \hline
$ {\rm Z}\rightarrow$ hadrons & 0.048          & 0.052          & 0.024          \\
$\eee$                        & 0.010          & 0.010          & 0.001          \\
$\emm$                        & 0.010          & 0.010          & 0.001          \\
Two-photon interactions       & 0.011          & 0.011          & 0.001          \\ 
Track definition              & 0.035          & 0.035          & 0.003          \\ 
Double track resolution       & 0.012          & 0.012          & 0.001          \\
Photon conversions            & 0.017          & 0.017          & 0.004          \\ 
Monte Carlo statistics        & 0.032          & 0.032          & 0.007   \\ \hline
Total                         & 0.073          & 0.076          & 0.026          \\
\hline
\end{tabular}
\caption{Systematic uncertainties in \% on the branching fractions resulting from the listed 
sources and their combined values.}
\label{table:sys}
\end{center}
\end{table}

%\pagebreak[4]
%\newpage
\clearpage

\end{document}